\documentclass[prd,12pt]{article}
\pdfoutput=1 

\usepackage{tikz}
\usepackage{amsmath,amssymb,graphicx,multirow,xspace,slashed,array,booktabs}
\usepackage[colorlinks=true,urlcolor=blue,anchorcolor=blue,citecolor=blue,filecolor=blue,linkcolor=blue,menucolor=blue,pagecolor=blue]{hyperref}
\usepackage[compress,numbers]{natbib}
\usepackage{subfigure, placeins}
\usepackage{booktabs}
\usepackage{blindtext, rotating}
\usepackage{afterpage}
\usepackage{enumitem}
\usepackage{ marvosym }
\usepackage{authblk} 
\usepackage{verbatim}
\usepackage{soul} 
\usepackage[normalem]{ulem}
\usepackage{pifont}
\usepackage{booktabs}

\usepackage{floatrow}
\newfloatcommand{capbtabbox}{table}[][\FBwidth]

\usepackage[font=footnotesize,labelfont=bf]{caption}

\usepackage{lineno}

\allowdisplaybreaks

\long\def\symbolfootnote[#1]#2{\begingroup%
\def\thefootnote{\fnsymbol{footnote}}\footnote[#1]{#2}\endgroup}

\renewcommand{\textfraction}{0}
\renewcommand{\topfraction}{0.95}

\allowdisplaybreaks

\makeatletter
	
	\@addtoreset{equation}{section}
\makeatother

\usepackage{color}
\usepackage[english]{babel}
\usepackage{hyperref}

\newcommand{\Slash}[1]{{\ooalign{\hfil#1\hfil\crcr\raise.167ex\hbox{/}}}}

\newcommand{\beq}{\begin{equation}}  \newcommand{\eeq}{\end{equation}}
\newcommand{\bef}{\begin{figure}}  \newcommand{\eef}{\end{figure}}
\newcommand{\bec}{\begin{center}}  \newcommand{\eec}{\end{center}}
  
\newcommand{\laq}[1]{\label{eq:#1}}  
\newcommand{\Eq}[1]{Eq.~(\ref{eq:#1})}

\newcommand{\eq}[1]{(\ref{eq:#1})}
\newcommand{\Sec}[1]{Sec.\ref{chap:#1}}

\newcommand{\vev}[1]{ \left\langle {#1} \right\rangle }

\newcommand{\lac}[1]{\label{chap:#1}}

\def\({\left(}
\def\){\right)}

\def\O{{O}}

\def\tr{\mathop{\rm tr}}

\newcommand{\AND}{~{\rm and}~}

\newcommand{\GEV}{ {\rm \, GeV} }

\def\a{\alpha}
\def\b{\beta}

\def\d{\delta}

\def\f{\phi}
\def\g{\gamma}

\def\k{\kappa}
\def\l{\lambda}
\def\m{\mu}

\def\r{\rho}

\def\x{\xi}

\def\D{\Delta}

\def\L{\Lambda}

\def\*{\dagger}

\begin{document}
\renewcommand\bibname{\Large References}

\newcommand{\newc}{\newcommand}
\newc{\gsim}{\lower.7ex\hbox{$\;\stackrel{\textstyle>}{\sim}\;$}}
\newc{\lsim}{\lower.7ex\hbox{$\;\stackrel{\textstyle<}{\sim}\;$}}
\newc{\gev}{\,{\rm GeV}}
\newc{\mev}{\,{\rm MeV}}
\newc{\ev}{\,{\rm eV}}
\newc{\kev}{\,{\rm keV}}
\newc{\tev}{\,{\rm TeV}}
\newc{\MHT}{$H_T^{\text{miss}}$}
\newc{\MET}{$\slashed{E}_T$}
\newc{\MTT}{$M_{T2}$}

\newcommand{\ifb}{\,\mathrm{fb}^{-1}}
\newcommand{\ipb}{\,\mathrm{pb}^{-1}}
\renewcommand*\descriptionlabel[1]{\hspace\labelsep\normalfont #1}

\def\ln{\mathop{\rm ln}}
\def\tr{\mathop{\rm tr}}
\def\Tr{\mathop{\rm Tr}}
\def\Im{\mathop{\rm Im}}
\def\Re{\mathop{\rm Re}}
\def\bR{\mathop{\bf R}}
\def\bC{\mathop{\bf C}}
\def\lie{\mathop{\hbox{\it\$}}} 
\newc{\mz}{M_Z}
\newc{\mpl}{M_*}
\newc{\mw}{m_{\rm weak}}
\newc{\nr}[1]{N^c_R{}_{#1}}

\renewcommand{\a}{\alpha}
\newcommand{\da}{{\dot \alpha}}
\renewcommand{\b}{\beta}
\newcommand{\db}{{\dot\beta}}
\newcommand{\dg}{{\dot\gamma}}
\renewcommand{\d}{\delta}
\newcommand{\dd}{{\dot\delta}}
\renewcommand{\r}{\rho}
\renewcommand{\l}{\lambda}
\renewcommand{\L}{\Lambda}
\renewcommand{\k}{\kappa}
\renewcommand{\th}{\theta}
\newcommand{\thb}{{\bar\theta}}
\newcommand{\B}{\bar B_\mu}
\newcommand{\cA}{c_{A_{u,d}}}
\newcommand{\cH}{c_{m_{u,d}}}
\renewcommand{\dag}{\dagger}
\newcommand{\Q}{\bar Q}
\renewcommand{\O}{O}

\newcommand{\CM}{{\mathcal M}}


\newcommand{\mhu}{{\hat m_{H_u}}}
\newcommand{\mhd}{{\hat m_{H_d}}}
\newcommand{\mhud}{{\hat m_{H_{u,d}}}}


\def\beq{\begin{equation}}
\def\eeq{\end{equation}}
\newcommand{\bea}{\begin{eqnarray}\begin{aligned}}
\newcommand{\eea}{\end{aligned}\end{eqnarray}}
\def\bitem{\begin{itemize}}
\def\eitem{\end{itemize}}

%
%

\renewcommand{\topfraction}{0.85}
\renewcommand{\textfraction}{0.1}
\renewcommand{\floatpagefraction}{0.75}



\renewcommand{\arraystretch}{1.3}

\baselineskip 0.6cm

\begin{titlepage}

\vspace*{-0.5cm}

\thispagestyle{empty}

\begin{center}
 \begin{flushright}
 TU-1195
 \end{flushright}
\vskip 1cm

{\LARGE \bf
Role of QCD in moduli stabilization during inflation and axion dark matter}

\vskip 1cm

\vskip 1.0cm
{\large Ryuichiro Kitano$^{1,2}$, Motoo Suzuki$^{1,3}$ and Wen Yin$^4$}
\vskip 1.0cm
{\it
$^1$ KEK Theory Center, Tsukuba 305-0801, Japan\\
$^2$Graduate University for Advanced Studies (Sokendai), Tsukuba 305-0801, Japan,
$^3$Department of Physics, Harvard University, Cambridge, MA, 02138, USA,\\
$^4$ Department of Physics, Tohoku University, Sendai, Miyagi 980-8578, Japan}
\vskip 1.0cm

\end{center}

\vskip 1cm

\begin{abstract}
Ignorance of the initial condition for the axion dynamics in the early
Universe has led us to consider an $O(1)$ valued initial amplitude,
and that prefers the decay constant, $F_a$, of the QCD axion to be an
intermediate scale such as $10^{12}$~GeV in order to explain the dark
matter abundance.
We explore a cosmological scenario of $F_a$ being
much larger than $10^{12}$~GeV by considering the axion and moduli dynamics during inflation to set the initial amplitude.
We show that if the volume moduli (radion) of the extra-dimension is
stabilized mainly by the QCD contribution to the moduli potential
during inflation, the QCD axion with the string-scale decay constant
obtains a mass around the inflationary Hubble parameter. This means
that the axion rolls down to the $\theta = 0$ minimum during the
inflation realizing almost vanishing initial amplitude, and the
inflationary quantum fluctuation can be the dominant source of the
current number density of axions. We find natural parameter regions
where the axion explains the cold dark matter of the Universe, while
the constraint on the isocurvature perturbation is avoided. The
presence of the axion miniclusters or axion stars are predicted in a
wide range of parameters, including the one explains the Subaru-HCS
microlensing event.

\end{abstract}

\flushbottom

\end{titlepage}


\section{Introduction}\label{introduction}

In string theory, the QCD axion, that solves the strong CP problem
~\cite{Weinberg:1977ma,Wilczek:1977pj,
Kim:1979if,Shifman:1979if,Zhitnitsky:1980tq,Dine:1981rt}, may
naturally arise via the compactification of
extra-dimensions~\cite{Witten:1984dg,
Svrcek:2006yi,Conlon:2006tq,Arvanitaki:2009fg,Acharya:2010zx,
Higaki:2011me, Cicoli:2012sz,Demirtas:2018akl,Marsh:2019bjr,
Mehta:2020kwu,Chen:2021hfq,Mehta:2021pwf,Cicoli:2022fzy}. (See
Refs.\,\cite{Jaeckel:2010ni,Ringwald:2012hr,Arias:2012az,Graham:2015ouw,Marsh:2015xka,Irastorza:2018dyq,
DiLuzio:2020wdo} for reviews.)  The axion decay constant, $F_a$, is
typically around the scale of the compactification. However, it was
known that the abundance would be too much if the compactification
scale is around the string scale $F_a = 10^{15-17}\,$GeV and if the
initial misalignment angle has the natural value of
$\O$(1)~\cite{Preskill:1982cy,Abbott:1982af,Dine:1982ah}. Even if we
tune the initial amplitude to be small, this is known to be in
conflict to high scale inflation because the generation of the
isocurvature
perturbation~\cite{Linde:1984ti,Linde:1985yf,Seckel:1985tj,Lyth:1989pb,Lyth:1991ub,Turner:1990uz,Linde:1991km}
constrains the inflation scale to be much lower than $H_{\rm inf}\sim
10^{13}$\,GeV from the CMB observation~\cite{Planck:2018jri}.

Recently, it was shown in Ref.~\cite{Kitano:2021fdl} (see also the
last part of the introduction for the earlier works) that the
abundance and isocurvature problems can be avoided by an enhanced
axion mass generated by small instantons.
In particular, in an extra-dimensional set-up,
one can naturally have a large QCD coupling 
during inflation due to the modified potential
for the volume moduli (radion).%
\footnote{We generically refer to the radion as the moduli in this paper.} 
This scenario is self-consistent since the realization of the axion
from compactification requires the stabilization of the size of the
extra-dimension both in the current Universe and in the inflationary
era. The moduli is generically light compared to the size of the
extra-dimension and it is mandatory to discuss its stabilization. 
In the context of supersymmetry (SUSY), a light QCD axion implies that
there should be a scalar superpartner, or string modulus with a mass
generated by the SUSY breaking, couples to the gluon.
The moduli stabilization in the perturbative regime, i.e., within the
effective theory, means that we need delicate balance among terms
which are in the different orders in the weak coupling expansion. For
example, in the case of the volume moduli, the different order terms
in the inverse volume expansion should balance to find a minimum.
Moreover, the cancellation of the cosmological constant in the current
Universe requires additional delicate balance among those term.
It is then natural to assume that this detail balance is badly broken
during the inflation where we have non-vanishing vacuum energy.

If the moduli have different values during inflation, there is a chance
that the QCD gets stronger during the inflation, and contribute 
significantly to the moduli potential as well as the axion potential.
In this paper, we find that if the radius of the extra-dimension is
stabilized between the Hubble-induced mass during inflation and {\it
the QCD induced potential}, the QCD axion mass is naturally around the
Hubble parameter during inflation, thanks to the fact that the up-type
Yukawa coupling $y_u$ in the Standard Model happens to satisfy $|y_u|
\sim (F_a/M_P)^2$. 
We also find that consistent with the high-scale inflation, the
inflationary fluctuations of the axion field can generate the correct
the DM abundance, without introducing the isocurvature problem due to
the time-varying axion mass.
We study the stabilization of the moduli in general set-up, and
show that the dangerous CP violating effects to regenerate a large
initial amplitude of the axion is naturally avoided due to the
symmetry of the  extra-dimension spacetime. 
\\

A heavy axion due to stronger QCD during the inflation was first
pointed out by Dvali \cite{Dvali:1995ce} in the context of alleviating
the axion abundance by using the moduli field dependent gluon
coupling. It was, however, soon pointed out that there are CP problems
in general for this solution to the over-production
problem~\cite{Banks:1996ea}.

For example, in the setup of the minimal SUSY standard model (MSSM), a
large field value of the Higgs field during inflation makes the quarks
heavier than those today and thus QCD gets stronger via the loop
effect. In that case, it was pointed out that generic CP phases of
soft breaking term spoil the solution to the over-production
problem~\cite{Choi:1996fs}. 
On the other hand, not as a solution to suppress the initial amplitude,
there have been a discussion of suppressing the isocurvature perturbation
by stronger QCD~\cite{Jeong:2013xta}. 
More recently, this idea was studied in \cite{Co:2018phi} by assuming
approximate CP symmetry in MSSM soft breaking parameter, while in
\cite{Ho:2019ayl, Matsui:2020wfx} (in the context of the stochastic
axion scenario mentioned below), a standard model setup has the CP violation 
accidentally absent. Compared to previous works we will show:
\begin{itemize}
\item in a relatively generic setup with an extra dimension, the
volume modulus can be stabilized during the inflation by a potential
whose main contribution is induced by the QCD dynamics. The heavier
QCD axion during inflation is then naturally realized with an
accidentally CP-safe structure thanks to the five-dimensional
spacetime. 
Namely, the CP violation is suppressed by the volume of the
extra dimension. 
\end{itemize}

In the paper \cite{Dvali:1995ce}, it was commented that the quantum
fluctuation of the axion field during inflation may play an important
role in explaining its abundance. However, now it is widely accepted
that such a scenario is excluded by the problem of too large
isocurvature perturbation. One possible way out is the scenario of the
stochastic axion~\cite{Graham:2018jyp,Takahashi:2018tdu}, in which the
small quantum fluctuation with $H_{\rm inf}\sim \Lambda_{\rm QCD}$,
accumulates to form an equilibrium state, which favors a
parametrically small initial axion misalignment explaining the axion
dark matter. This scenario requires a moderately low inflation scale
and a very long inflation period for the fluctuation to accumulate.
Compared to the previous work, we show that
\begin{itemize}
\item axion dark matter from the large quantum fluctuation can be
realized if the heavy axion becomes lighter than the Hubble parameter
during the last few e-folds of the inflation. In contrast to the
stochastic axion scenario, the axion fluctuation is produced in a very
short period with a large inflation scale. The mode evolution is very
different, and the mechanism may be probed from the observation of
tensor-to-scalar mode and further measurement of the dark matter
isocurvature. Miniclusters and axion stars can be formed due to the
dominant fluctuation in the intermediate scale.  
\end{itemize}

The organization of the paper is as follows. In Sec.~\ref{sec:revisit},
we revisit the scenario of stronger QCD and relate it to the moduli
stabilization during inflation. In Sec.~\ref{sec:abundance}, we discuss the axion dark
matter by especially focusing the new scenario of the production by
isocurvature-safe axion fluctuation and discuss the minicluster
formations. In Sec.~\ref{sec:model} and \Sec{UVMODEL}, we construct
models for the moduli stabilization with QCD induced potential during
inflation. The final section is devoted to discussion and conclusions.

\section{Moduli and axion stabilizations by stronger QCD during inflation}
\label{sec:revisit}

In this section, we describe the main idea of the stabilization of the
moduli and axion fields by the non-perturbative QCD effects during
inflation. It turns out that both fields are simultaneously stabilized
in the early stage of inflation and the axion mass is comparable to
the Hubble scale. 

We consider the cosmological scenario where a modulus field, $T$,%
\footnote{In precise, the moduli is described as $\Re T$ in Sec.~\ref{chap:UVMODEL}.}
has
different field values during and after inflation due to the modified
potential shape such as by the Hubble induced mass term. In
supergravity and superstring theories, a kinetic term of a gauge field
in the Standard Model is obtained via 
\begin{align}
   \mathcal{L}=f\left(T\right)\, F_{\mu\nu} F^{\mu\nu}\ ,
\end{align}
where $f(X)$
denotes a function of $X$, and $F_{\mu\nu}$ denote a gauge field
strength tensor.  
$f(\vev{T})=1/(4g^2)$
with $g$ is the measured gauge coupling in
the present vacuum. $\vev{X}$ denotes the expectation value of $X$. However, $g$ may be different in the early Universe depending on $\langle T\rangle_{\rm inf}$, where we use sub/super script of $\rm inf$ to denote the value during inflation, here and hereafter.

In addition, an axion, $a$, couples to the gauge field,
\beq 
\delta \mathcal{L}=\frac{a}{32\pi^2 F_a}  F_{\mu\nu} \tilde F^{\mu\nu}\ ,
\eeq 
where $\tilde F_{\m\nu}$ denotes the dual of $F_{\m\nu}$, which may be either abelian or non-abelian, in which case we omit the gauge index. In SUSY
extensions, the axion is regarded as the superpartner of a modulus, while
we do not restrict ourselves to SUSY models. In the following, we will
always assume that the (perhaps time-varying) masses of SUSY partners
except for the modulus are well above the energy scale, e.g. the
Hubble parameter, under consideration for simplicity of discussion.
Alternatively, the axion may also originate from the extra-dimensional
Wilson loop of the dark gauge field. In these cases, a natural scale
of the axion decay constant is the compactification scale, that is
naturally,
\beq 
F_a =10^{15-17}\GEV\ .
\eeq

We are interested in the QCD effects to the moduli potential during
inflation.
Indeed, if the $SU(3)_c$ gauge coupling during inflation is larger
than the current one, the QCD confinement scale, $\Lambda_{\rm QCD}$,
become much larger than the ordinary QCD scale $\sim 100$\,MeV.
The axion mass is obtained due to the non-perturbative effect, both
during inflation and after inflation,
\begin{align}
\label{eq:axion_mass}
& m_a^2\sim \frac{m_u (\Lambda_{\rm QCD})^3}{F_a^2}\ ,
\end{align}
where $m_a$ denotes the axion mass, and $m_u$ is the up-quark mass. 
This form should be valid as long as $m_u$ is smaller than
$\Lambda_{\rm QCD}$ and it is the lightest quark. Namely, it does not
depend on how many quarks are lighter than the QCD scale. It should be noted that such
a quantity may depend on cosmic time during inflation.

In~\eqref{eq:axion_mass}, the up-quark mass is
\begin{align}
    m_{u,\rm inf}\sim y_u \langle h\rangle_{\rm inf}\ , 
\end{align}
where $\langle h\rangle_{\rm inf} $ is the expectation value of the SM Higgs
field during inflation, and $y_u\sim 10^{-5}$ is the up-type Yukawa
coupling.
When $\Lambda_{\rm QCD}^{\rm inf}$ is high enough and the chiral
symmetry is broken in the 6 flavor QCD with all the SM quarks
``light", the QCD itself can induce an electroweak symmetry breaking
by the top quark condensation that gives a tad pole term of the Higgs
field,
\beq 
 \vev{h}_{\rm inf}\sim \Lambda_{\rm QCD}^{\rm inf}\ .
\eeq
Here, the self-coupling of the Higgs field and the top quark Yukawa
coupling are assumed to be of order unity. The axion mass during
inflation is 
\beq 
\laq{mainf}
m_{a,\rm inf}^2 \sim y_u\frac{(\Lambda_{\rm QCD}^{\rm inf})^4}{F_a^2}
\ .\ 
\eeq
Here, for simplicity, we assume that $F_a$ does not depend on time, which will be revisited when we discuss concrete models. 
We note that the Higgs field should not acquire a much larger
(positive or negative) mass squared, $(M_h^{\rm inf})^2$, during
inflation than the inflationary QCD scale. 
As we will see, the QCD scale during inflation is comparable to the
inflation potential scale in order to stabilize the modulus. If the
$|(M_{h}^{\rm inf})^2|$ were too large, it would contribute to the
inflation potential via Coleman-Weinberg corrections, which usually
spoil the inflation dynamics, or, at least, it requires a fine-tuning
among contributions to the total potential. 
We also note that the Higgs field may be driven to be very large value 
around the Planck scale during inflation in SUSY
models~\cite{Choi:1996fs}, while it usually does not work well without
a fine-tuning of the CP phase~\cite{Choi:1996fs} (however, the model
in \Sec{UVMODEL} may work%
\footnote{The Higgs VEV during inflation can be much larger than $\Lambda_{\rm QCD}^{\rm inf}$ due to the coupling with $e.g.$ the inflaton field~\cite{Matsui:2020wfx}, and the lightest quark mass (up-quark mass) becomes larger than $\Lambda_{\rm QCD}^{\rm inf}$. Then, the axion mass is given by
$m_{a,\rm inf}^2\sim \frac{(\Lambda_{\rm QCD}^{\rm inf})^4}{F_a^2}~~~(m_u>\Lambda^{\rm inf}_{\rm QCD})$. In this case, the coincidence found in this paper does not apply.  
The condition $m_a>H_{\rm inf}$ is satisfied in a broader parameter space because the axion mass is not suppressed by the factor of $y_u$. We need an additional CPV source for explaining dark matter abundance since it is then natural to have the axion mass to be always heavier than the Hubble parameter during inflation. 
}). Although we have in mind that SUSY is in some high energy scale, we do not consider the possibility that the some MSSM field excursion is so large that it introduces a misalignment of the QCD axion potential minima during and after the inflation.

Now let us consider the moduli stabilization during inflation by
taking into account of the back reaction from the QCD dynamics, which
is assumed to be a dominant source for the stabilization. If the QCD
contribution is dominant, we expect that its contribution to the
moduli mass is comparable to the Hubble induced mass, namely
the Hubble scale,
\beq 
m_{\rm moduli, inf}^2\sim \frac{(\Lambda_{\rm QCD}^{\rm inf})^4}{M_P^2}\sim H_{\rm inf}^2.
\eeq 
with $m_{\rm moduli, inf}$, $H_{\rm inf}$ being the mass scale of the
moduli and the Hubble scale during inflation, respectively. 
For the axion mass during inflation, we find 
\beq 
m_{a,\rm inf}^2\sim \frac{y_u \(\L^{\rm inf}_{\rm QCD}\)^4}{F_a^2} \sim \frac{y_u M_P^2}{F_a^2}\times \frac{\(\L^{\rm inf}_{\rm QCD}\)^4}{M_P^2}.
\eeq
With $F_a\sim 10^{15}\GEV$ during inflation, it is natural that the first factor is order 1 and therefore we find 
\beq 
\laq{maHinf}
\boxed{\Rightarrow m_{a,\rm inf}^2 \sim H_{\rm inf}^2. }
\eeq 
As a consequence, we notice that if the QCD potential plays an
important role in stabilizing the moduli, the axion mass during
inflation is comparable to the Hubble parameter. 

Motivated by this observation, in the following we study axion
productions when \Eq{maHinf} is satisfied. The initial amplitude is
almost zero in this case as the axion is settled into its minimum during inflation, while it is also sensitive to the inflationary fluctuations.
We also discuss the moduli stabilization during inflation in a
realistic set-up.

\section{Dark matter axion from isocurvature fluctuations}
\label{sec:abundance}

\subsection{Axion fluctuation during inflation}

The axion is stabilized around its potential minimum when $m_{a,{\rm
inf}}$ is large enough compared to the Hubble constant $H_{\rm inf}$
during inflation, 
\begin{align}
\label{eq:ma_HI}
m_{a,{\rm inf}} \gtrsim  H_{\rm inf}\ .
\end{align}
We consider the case where this is satisfied at the moment of the
horizon exit of the scale of the cosmic-microwave background (CMB). If
there is no new source of CP violation in the axion potential, which
is the assumption to be justified, the initial misalignment angle of
the axion is approximately zero, $i.e.$ the axion potential minimum is
the same as that at low temperatures\footnote{In more detail, even
without new CP sources, there is an enhanced contribution from the CKM
phase which may disturb the axion potential
minimum~\cite{Linde:1996cx}. This effect is proportional to the
Jarlskog invariant $J$, $i.e.$ ${\mit \Delta}\theta= c_J J$ with
$c_J\lesssim 1$ where ${\mit \Delta}\theta$ denotes the difference of
the axion potential minimum from that at the present minimum. Unless
the Jarlskog invariant during inflation is much larger than the
present one $J\sim 10^{-5}$, its contribution to the misalignment
angle of the axion dark matter is negligible.}.

The absence of new CP violating source is a crucial assumption in
order for the minimum of the potential during inflation to be zero.
Potentially dangerous possibilities are CP violating configurations of
the moduli/inflaton during inflation and the presence of CP violating
interactions which become important when the Higgs VEV is large
and/or QCD coupling is strong.
The moduli/inflaton configurations can easily be CP conserving by
assuming that the moduli and inflaton are both CP even fields. For
example, if the moduli we consider is the moduli of an
extra-dimension, it is CP even as it is a component of the metric. We
discuss this possibility in more detail in \Sec{UVMODEL}. The
enhancement of the small instanton effects together with the higher
dimensional operators generically give a large CP violating
contribution to the axion potential~\cite{Kitano:2021fdl}.
This contribution, however, depends heavily on the particle content of
the theory as well as the evolution of the coupling constant
at high energies. The contribution can be negligible, especially, with a large (but not too large) extra-dimension (and SUSY) we consider later in this paper.

Under those assumptions, the axion abundance from the usual misalignment or realignment
mechanism is almost zero which means the upper bound on $F_a$
 disappears if there were no other contributions.
In the following, we discuss the contribution to the axion abundance
from the quantum fluctuation of the axion field during inflation. We
will see that this contribution can be sizable enough to explain dark
matter of the Universe without contradicting the isocurvature
constraints from the CMB observations.

The QCD confinement scale during inflation changes from moment to
moment because the moduli field may also be rolling through the change
of its potential, and because the QCD scale is sensitive to the moduli
field in an exponential way. Less than $O(10\%)$ change of the modulus
field value would alter the QCD scale extremely because the large
coefficient of the beta function of the gauge coupling. Thus, even if
$m_{a,{\rm inf}}\gtrsim H_{\rm inf}$ is satisfied at an early stage of
the inflation, $m_{a,{\rm inf}}$ may become smaller than $H_{\rm inf}$
before the end of inflation.
The constraint from the isocurvature fluctuation can be avoided if the
time of $m_{a,{\rm inf}}$ being smaller than $H_{\rm inf}$ happens
sometime after when the inflaton fluctuation of the CMB scale exits the horizon.

In general the axion field can be decomposed by 
\beq 
a= a_{\rm 0}+ \delta a 
\eeq 
with $a_{\rm 0}$ being the homogeneous mode in the observable
Universe, which is set to zero due to \Eq{ma_HI} around the horizon exit of the CMB scale,
\beq
a_0=0.
\eeq
Once the axion becomes lighter than the Hubble parameter, the axion
field fluctuation, $\d a$, gets of the order $\delta a\sim H_{\rm
inf}/2\pi$. We define $k_{\rm cutoff}$ as the comoving momentum of the
axion fluctuation 
when the axion mass becomes comparable to the Hubble parameter during
inflation, $i.e.$ $m_a=H_{\rm inf}$, and $k_{\rm cutoff}$ denotes the
smallest comoving momentum with the axion fluctuation.%
\footnote{The relation between the e-fold and the comoving momentum
$k$ is given as $N\approx 54.4+\ln\left(\frac{10^{-3}\,{\rm
Mpc}^{-1}}{k}\right)+\frac{1}{3}\ln\left(\frac{T_R}{10^{10}\,{\rm
GeV}}\right)+\frac{1}{3}\ln\left(\frac{H_{\rm inf}}{10^{10}}\,{\rm
GeV}\right)$ where the scale factor is taken as $R=1$ at present. }

We use the power spectrum with the cutoff of $k_{\rm cutoff}$: 
\beq\laq{PS}
\vev{\d a_k \d a_{-k}}\sim \frac{H_{\rm inf}^2}{(2\pi)^2k^3} \Theta(k-k_{\rm cutoff})\ ,
\eeq 
where $\delta a_k$ is the Fourier transform of the canonically
normalized axion field, $\Theta$ is the heaviside step function, and
$k$ denotes the comoving momentum. Note that the spectrum around the
cutoff may depend on the detail of the moduli potential, and how the
axion mass becomes smaller than the Hubble parameter, while the
behavior with $k\ll k_{\rm cutoff}$ after the inflation, where the
axion mass is neglected, is almost model-independent. It represents
the well-known scale-invariant power spectrum in the de-Sitter space
in the wave number range. The fluctuation is frozen until it reenters
the horizon or the axion mass becomes important. The fluctuation
 characterized with the power-spectrum \Eq{PS} is considered as the
initial condition set at the end of inflation.

\subsection{Axion fluctuation after inflation}
\label{sec:fluc_after_inf}
For simplicity, we consider the case where the Universe after
inflation is soon dominated by radiation (the extension with a
matter-dominated-reheating era is straightforward). The
fluctuation evolves via the equation of motion after the end of
inflation, 
\beq
\delta \ddot a_k(t)+ 3H \delta\dot a_k(t)
+\left(\frac{k^2}{R^2}+m_a(t)^2\right) \d a_k(t) =0 
\laq{flueom}
\eeq
where $R$ is the scale factor, and $m_a(t)$ denotes the time dependent
axion mass. 
In the regime of $(k/R)^2+m_a^2 \ll H^2$, we can neglect
the gradient contribution and $\d a_k$ is frozen at the value soon
after inflation.

The mode $k$ starts to oscillate when $(k/R)^2+m_a^2\gg H^2$ gets
satisfied. By noting that $k/R \AND m_a$ change slowly, the
differential ``number density" starts to decrease due to the red
shift,
\beq 
n_{a, k}(t)= (\sqrt{\frac{k^2}{R(t)^2}+m_a(t)^2} |\d a_k(t)|^2)\propto R(t)^{-3}\ .
\eeq 
If the mode $k_{\rm cutoff}$ reenters the horizon much before the quark/hadron transition, $i.e.$ if $k_{\rm cutoff}/R =H$ is satisfied at $H\gg m_a$, the number density to entropy ratio for each $k$ mode is estimated as,
\beq 
\frac{n_a^{\rm fluc}}{s}[k] = 
k^3 \left.\,\frac{n_{a,k}}{s}\right|_{H=k/R}\sim \left.
\frac{H H_{\rm inf}^2}{s (2\pi)^2}\right|_{H=k/R}\ .
\eeq 
The axion abundance is 
\beq 
\Omega_{a}^{\rm fluc}=\frac{s_0}{\rho_c} m_a^0 \int_{k_{\rm cutoff}}^{R_{\rm inf} H_{\rm inf}}d\ln k'\,\frac{n_a^{\rm fluc}}{s}[k']\sim 10^{-3}\(\frac{\sqrt{g_\star}/g_{\star ,s }}{0.1}\)\frac{10^{15}\GEV}{F_a}\(\frac{H_{\rm inf}}{10^{13}\GEV}\)^2 \frac{1\GEV}{T_{\rm cutoff}}
\eeq 
where $s_0~(\rho_c)$ is the present entropy density (critical
density), $g_\star~(g_{\star,s})$ is the relativistic degrees of
freedom for energy (entropy) density, and $m_a^0\equiv m_a[t\to
\infty]$ is the axion mass in the vacuum. We have defined  $T_{\rm
cutoff}$ as the temperature $H=k_{\rm cutoff}/R$. A remarkable
difference from the standard misalignment mechanism is that the above
form of the abundance depends on the decay constant inversely. This is
because neither the ``initial amplitude" nor the cosmic time for the
onset of oscillation depends on the axion mass. Only when we convert
the number density into the energy density the axion mass enters in
the estimation, giving the inverse dependence of $F_a$.

On the other hand, when 
the mode of $k_{\rm cutoff}$ reenters the horizon much after the quark/hadron transition, $i.e.$
$k_{\rm cutoff}/R =H$ happens at $H\ll m_a$, the dominant component is from a production similar to the misalignment mechanism. 
The difference is that we have a cutoff in the IR modes, which will be important to alleviate the isocurvature bound. We call the production in this region, quantum misalignment mechanism.
By defining the mode $k_{a}=m_a R_a$ at $H=m_a$, when the scale factor is $R_a$, all the modes in the range $k_{\rm cutoff}<k< k_a$ can be regarded as ``zero'' mode and they start to oscillate at the moment according to \Eq{flueom}. 
The number density to entropy density ratio from this component is evaluated as 
\beq 
\frac{n_a^{\rm qmis}}{s}=\left.\frac{1}{s}\right|_{H=m_a} \times \int_{k_{\rm cutoff}}^{k_a} dk k^2{n_{a,k}}\approx \left.\frac{H H_{\rm inf}^2}{(2\pi)^2 s}\right|_{H=m_a} \times \log(\frac{k_a}{k_{\rm cutoff}}). 
\eeq 
If this component dominates, 
by redefining the misalignment angle $\theta_i= \frac{H_{\rm inf}}{2\pi F_a}\sqrt{\log(k_a/k_{\rm cutoff})}$
we have the abundance from the usual formula~\cite{Ballesteros:2016xej},\footnote{The estimation of the abundance depends on the temperature dependence of the topological susceptibility~\cite{Abbott:1982af,Preskill:1982cy,Dine:1982ah}. See~\cite{Berkowitz:2015aua,Kitano:2015fla,Borsanyi:2015cka,Petreczky:2016vrs,Frison:2016vuc,Borsanyi:2016ksw} for the recent lattice results.
A recent estimation of the abundance is given in~\cite{Ballesteros:2016xej}.
}
\begin{align}
\label{eq:omega_a}
\Omega_a^{\rm qmis} h^2=0.2\(\frac{H_{\rm inf}}{2\pi F_a} \)^2
{\mit\Delta N}
\left(\frac{F_a}{10^{12}{\rm GeV}}\right)^{1.19}\ ,
\end{align}
where ${\mit\Delta N}$ is defined as
\beq 
{\mit\Delta N}\equiv \log(\frac{k_a}{k_{\rm cutoff}})\ .
\eeq

In summary, the axion abundance is given in two regimes,
\beq 
\label{eq:omegaa}
\Omega_a= \frac{s_0}{\rho_c} m_a^0 \times \begin{cases}
\displaystyle{
\int_{k_{\rm cutoff}}^{R_{\rm inf} H_{\rm inf}}d\ln k'\,\frac{n_a^{\rm fluc}}{s}[k'] \vspace{3mm} ~~~~~~~~~~~~~~~~~~~~~~~(~\text{if} ~k_{\rm cutoff}\gg k_a)}\\
\displaystyle{\frac{n_a^{\rm qmis}}{s}+\int_{k_a}^{R_{\rm inf} H_{\rm inf}}d\ln k'\frac{n_a^{\rm fluc}}{s}[k']~~~~~~~~~~~~~~~(~\text{if} ~k_{\rm cutoff}\ll k_a)}\\
\end{cases},
\eeq 
It should be noted that a simple power spectrum in~\eq{PS} is assumed in our estimation as mentioned before, but the detail distribution around the mode $k_{\rm cutoff}$ is model-dependent and it may affect the final abundance estimation.
Nonetheless, we shall use the estimation in~\eqref{eq:omegaa} in the following discussion.

\begin{figure}
\centering
\hspace{1cm}
\includegraphics[width=0.6\linewidth]{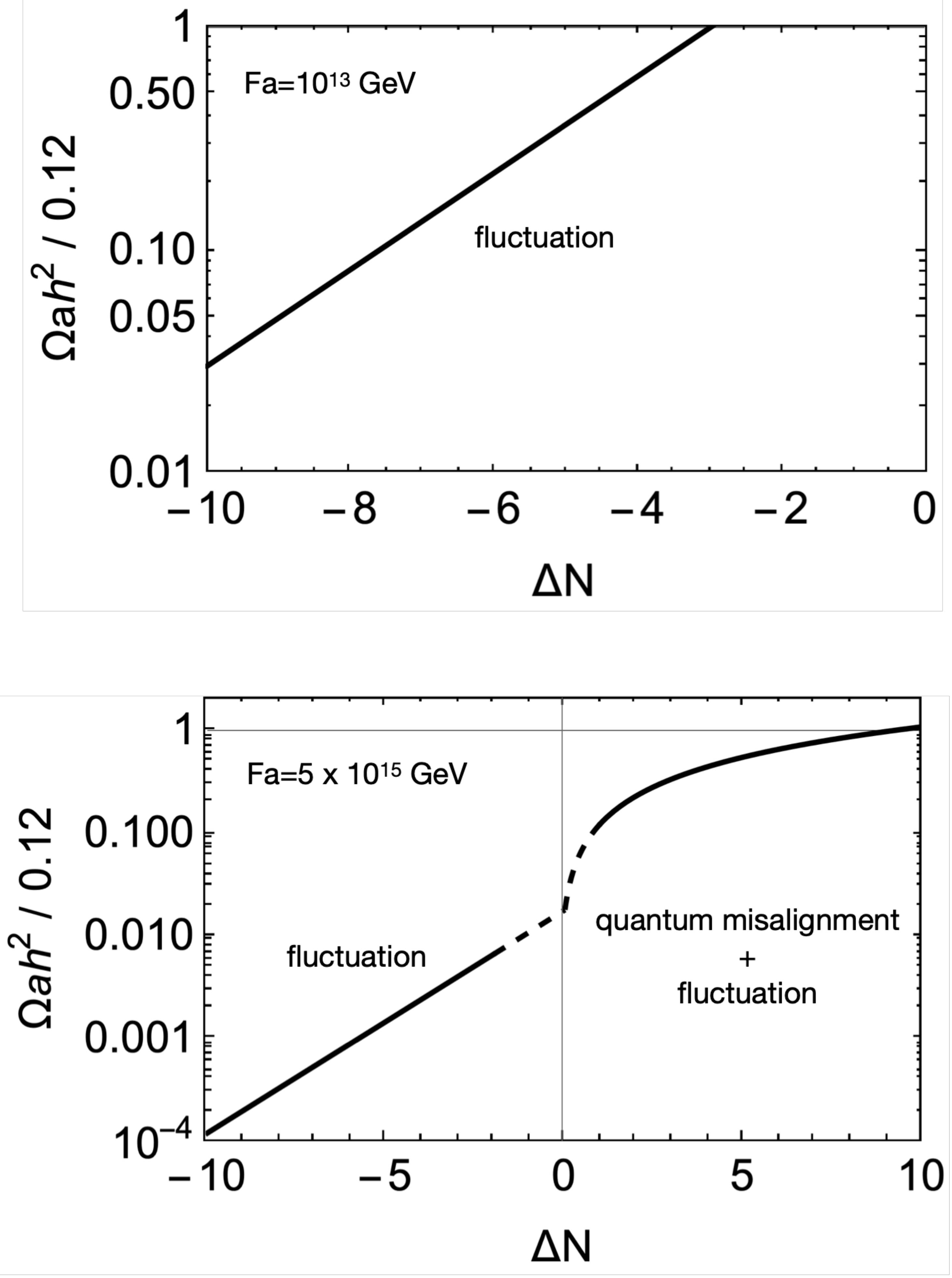}
   \vspace{0.2cm}
   \caption{The axion dark matter abundance as a function of ${\mit \Delta}N$ (black solid line). 
   In the upper figure, we took $F_a=10^{13}\,{\rm GeV}$ and  $H_{\rm inf}=5\times 10^{13}\,{\rm GeV}$.
   In the lower figure, we took $F_a=5\times 10^{15}\,{\rm GeV}$ and  $H_{\rm inf}=5\times 10^{13}\,{\rm GeV}$.
   Uncertainty of the estimation of the dark matter abundance around ${\mit \Delta}N$ ($k_{\rm cutoff}\sim k_a$) is denoted by black dashed lines.  }
\label{fig:dN_Oh2}
\end{figure}

Fig.~\ref{fig:dN_Oh2} shows the abundance of the axion dark matter by varying $k_{\rm cutoff}$ or ${\mit\Delta N}$ ($k_{\rm max}$) (black solid lines). 
In the upper figure,  we took $F_a=10^{13}\,{\rm GeV}$ and  $H_{\rm inf}=5\times 10^{13}\,{\rm GeV}$ for which the effective theory description is barely satisfied by noting the Gibbons Hawking temperature is $H_{\rm inf}/2\pi$ and the cutoff scale is $2\pi F_a$. 
The current total dark matter abundance is explained only by the high momentum modes of $k\geq k_{\rm cutoff}\gg k_a$.
In the lower figure, we took $F_a=5\times 10^{15}\,{\rm GeV}$ and  $H_{\rm inf}=5\times 10^{13}\,{\rm GeV}$.
The black dashed line denote uncertainties of our estimation around ${\mit \Delta}N\sim 0$ ($k_{\rm cutoff}\sim k_a$). 

\paragraph{Isocurvature problem}
In the regime of $\log(\frac{k_a}{k_{\rm cutoff}})\gg 0$, we have to care about the isocurvature bound. Indeed, if $k_{\rm cutoff}$ is too small, 
we encounter the isocurvature problem because our scenario approaches to the usual pre-inflationary PQ breaking scenario of the QCD axion.    
In addition to the CMB measurement, the isocurvature perturbation is also constrained by the Lyman $\alpha$ data because the matter power spectrum is distorted~\cite{Beltran:2005gr,Redi:2022llj}. To avoid these constraints, we require that a large amount of the axion isocurvature perturbation is only generated for the scale much below $\mathcal{O}(0.1)$ Mpc. 
This reduces to the condition ${\mit \Delta N}\lesssim 20$. 
This implies a lower bound $k_{\rm cutoff}\gtrsim 3\,{\rm Mpc}^{-1}$
, where we have defined that $R=1$ in the present Universe.

\begin{figure}
\centering
\hspace{1cm}
   \includegraphics[width=0.6\linewidth]{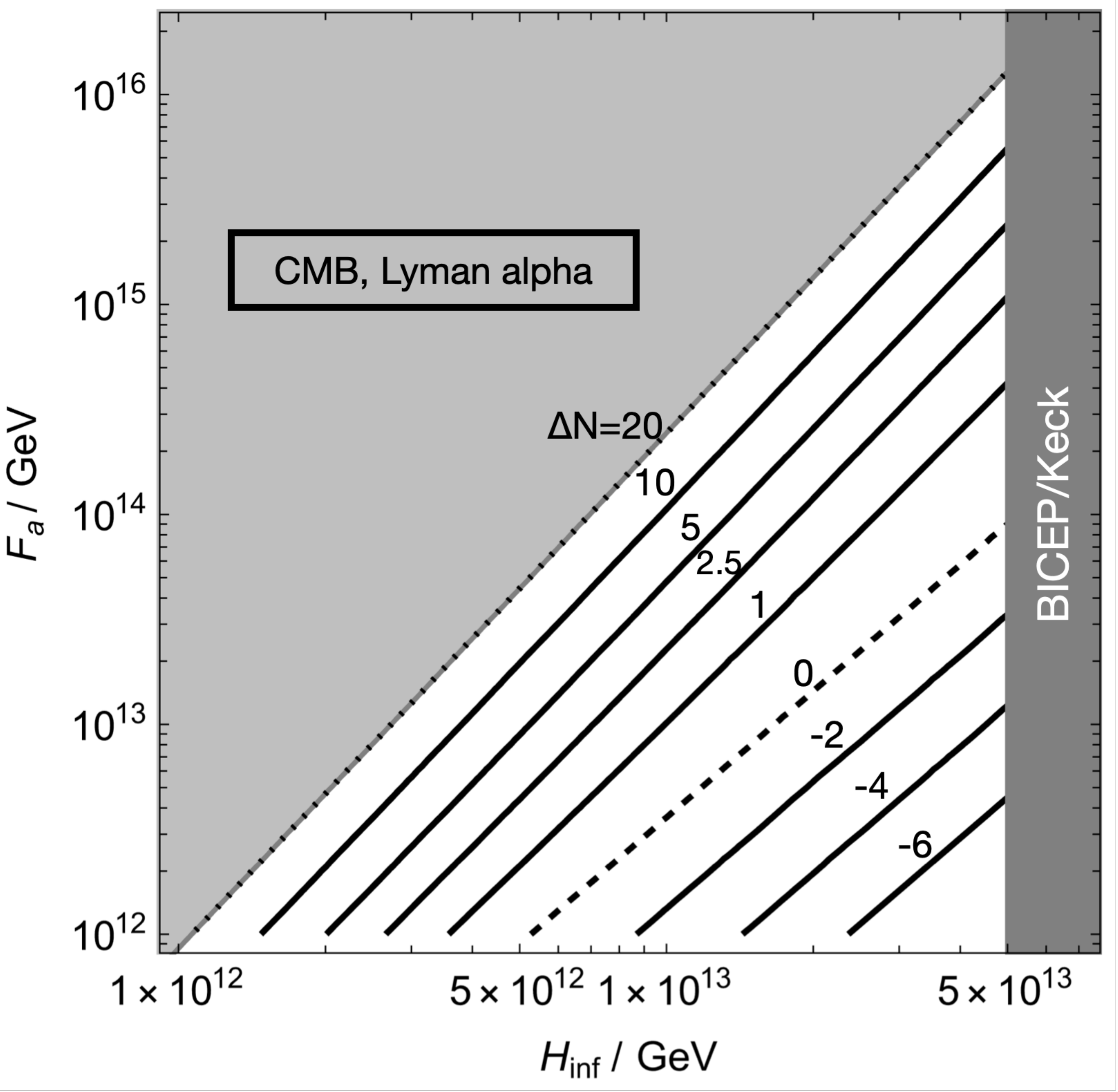}
   \vspace{0.2cm}
   \caption{The parameter region for QCD axion dark matter from inflationary fluctuations in $H_{\rm inf}-F_a$ plane.
   The black solid lines denote $\Omega_a h^2/0.12=1$ for the case that axion becomes light during inflation at ${\mit \Delta N}=-6,-4,-2,~0,~1,~2.5,~5,~10,~20$ from upper left  to lower right, respectively. The darker gray shaded region is the constraint from the BICEP/Keck. The light gray shaded region or more precisely ${\mit\Delta} N\gtrsim 20$ is constrained from the Lyman-alpha and CMB results.}
\label{fig:HI_Fa}
\end{figure}

The isocurvature perturbations at smaller scale can be also probed by the CMB spectral distortions~\cite{Zeldovich:1969ff,Sunyaev:1970eu,Hu:1992dc} which are the deviations from the black body spectrum of the CMB frequency spectrum.%
\footnote{See $e.g.$~\cite{Chluba:2014sma} for a brief overview of science with the spectral distortions.}
The distortions have the information before the recombination, and it can probe small-scale isocurvature perturbations~\cite{Chluba:2013dna}.%
\footnote{While the analysis assumes the simple power law spectrum, our scenario gives more complicated spectrum, in particular, when $k\sim k_a$.
} 
The current bound on the amplitude of the power spectrum of the CDM isocurvature mode has been studied in~\cite{Chluba:2013dna} by using the limits from COBE/FIRAS, and $O(1)$ amplitude is still allowed for $k\gtrsim 1\,{\rm Mpc}^{-1}$ if $e.g.$ the scale invariant power spectrum 
 is assumed.
Further probes of the distortions are prospected by several proposals, $e.g.$ PIXIE~\cite{Kogut:2011xw}, PRISM~\cite{PRISM:2013fvg}, 
Super-PIXIE~\cite{Kogut:2019vqh}, Voyage 2050~\cite{Chluba:2019nxa}, BISOU~\cite{Maffei:2021xur}.
For example, \cite{Chluba:2013dna} shows that PIXIE-type experiments can detect $\mathcal{O}(1)$ amplitude for 
the scale-invariant power spectrum.

In Fig.~\ref{fig:HI_Fa}, we have shown the current abundance of the axion dark matter on $H_{\rm inf}-F_a$ plane. 
The black solid lines denote $\Omega_a h^2/0.12=1$ for ${\mit \Delta N}=-6,-4,-2,~1,~2.5,~5,~10,~20$ from upper left  to lower right, respectively.
The darker gray shaded region denotes the bound from the BICEP/Keck Observations~\cite{BICEP:2021xfz}.
The lighter gray region denotes ${\mit\Delta N}>20$, which is constrained from the Lyman-alpha and the CMB measurements as discussed before.
We also note that the perturbativity condition of $m_a\lesssim F_a$ is satisfied in the figure.
Usually satisfying the bound of the isocurvature and explaining the axion abundance by the fluctuation requires a fine-tuning of the (time-dependent) potential so that the axion becomes light after the horizon exit of the CMB scale and before the end of inflation. We emphasize again in our scenario that this is as natural as the parameter region that axion remains always light or heavy during inflation thanks to \Eq{maHinf}.

\begin{figure}
\centering
\hspace{1cm}
\includegraphics[width=0.8\linewidth]{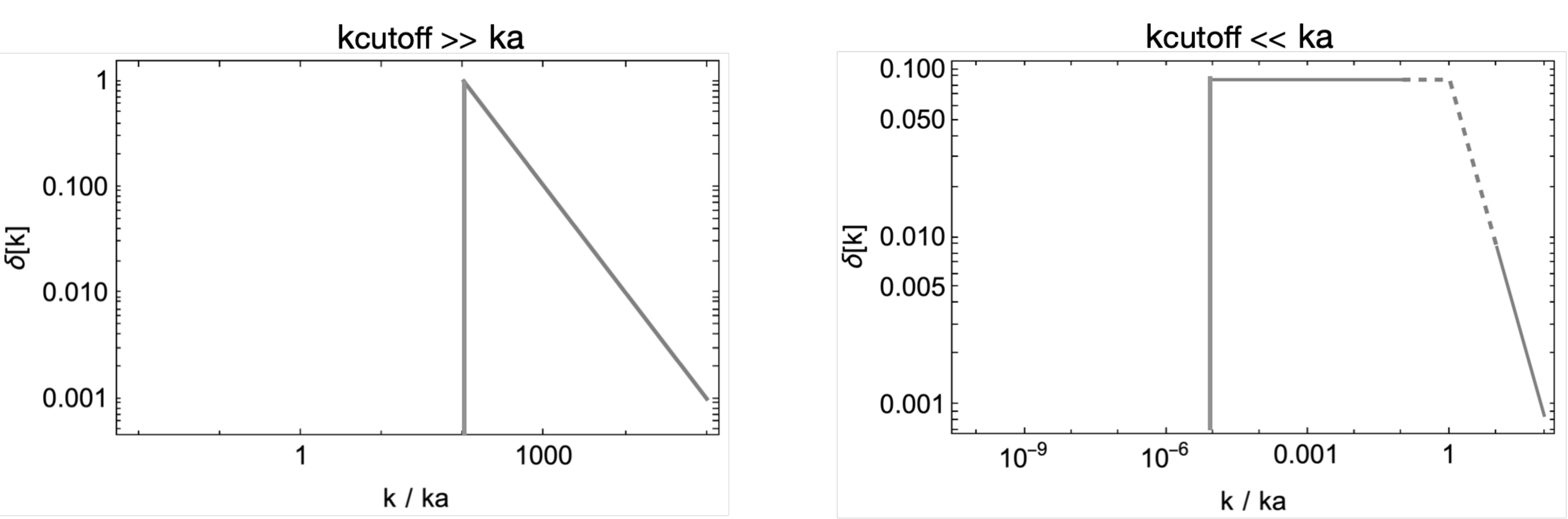}
   \vspace{0.2cm}
   \caption{$\delta[k]$ for $k_{\rm cutoff}\gg k_a$ (left) and $k_{\rm cutoff}\ll k_a$ (right).   We plotted $\delta[k]$ around $k=k_a$ by the dashed gray line due to the uncertainty of $n_a^{\rm fluc}[k]$ as discussed around Sec.~\ref{sec:fluc_after_inf}. 
   }
\label{fig:k_delk}
\end{figure}

\paragraph{Axion clumps}
Given that we have large amplitudes for small scales, 
there can be over-dense regions of axions at later times; the ratio of the fluctuation to the total dark matter energy density, parameterized by $\delta$, can be of order 1. More concretely, after when the $k_{\rm cutoff}$ mode starts to oscillate, 
the fluctuation for each mode, $\d[k]$, is obtained
from \Eq{flueom} as
\begin{align}
&\delta[k]\equiv \frac{\d_{\log k} \rho_a}{\rho_a}\\
&\sim \begin{cases}
\displaystyle{\frac{k_{\rm cutoff}}{k} \Theta[k-k_{\rm cutoff}]\times \Theta[H_{\rm inf}R_{\rm inf}-k] \vspace{3mm} ~~~~~~~(~\text{if} ~k_{\rm cutoff}\gg k_a)}\\
\displaystyle{\Theta[k-k_{\rm cutoff}]\(\frac{1}{\log{k_a/k_{\rm cutoff}}}\Theta[k_a-k] +\frac{k_a}{k\log{k_a/k_{\rm cutoff}}} \Theta[k-k_a]\Theta[H_{\rm inf}R_{\rm inf}-k]\)
(~\text{if} ~k_{\rm cutoff}\ll k_a) }\ .
\end{cases}
\end{align}
where $\d_{\log k} \rho_a\equiv k^3\times m_a \, n_{a,k}$  and we neglect the component without the logarithmic enhancement in $\rho_a$ for the case $k_a \gg k_{\rm cutoff}$.%
\footnote{For $k\gg k_a$, we have used $(k^3 n_{a,k})/s \propto H_{\rm inf}^2 (k/R_{\rm osc})H^{-3/2}$ with $k/R_k=H\propto R_k^{-2}$ in the radiation dominated Universe, where we defined $R_k$ as the scale factor when the oscillation starts for the $k$ mode. Since $R_k\propto k^{-1}$, we obtain $(k^3 n_{a,k})/s \propto k^{-1}$.}
We plotted $\delta[k]$ in Fig.~\ref{fig:k_delk}.

Such a density perturbation can be a seed for the axion minicluster/star or axiton formations as in the case of the axions from topological defects~\cite{Hogan:1988mp}~\cite{Kolb:1993zz,Kolb:1993hw,Kolb:1994fi,Kolb:1995bu} (see Refs. \cite{Vaquero:2018tib,Buschmann:2019icd,Ellis:2020gtq,Eggemeier:2019khm,Ellis:2022grh} for more recent simulations). 
The difference is the distribution of $\d[k]$ which has the overdensity in a wider range of $k$, especially for the case $k_{\rm cutoff}\ll k_a$, where $\d \sim 1$ is satisfied in a scale independent way in the range $k_{\rm cutoff}<k<k_a$. 
In addition, as we will see, our scenario does not have the stringent bound from the axion decay constant c.f.~Ref.\,\cite{Schiappacasse:2021zlr}.

While a detailed study is beyond our scope in this paper,
those axion clumps are possible to be explored by such as lensing~\cite{Zurek:2006sy,Hardy:2016mns,Fairbairn:2017dmf,Fairbairn:2017sil}, and even may explain the one indicated in the Subaru HSC microlensing~\cite{Niikura:2017zjd,Sugiyama:2021xqg}.
In addition, the axion search experiments can be also have distinguished behavior, because axionic dense clumps going through the detector can enhance the signal rate by many orders of magnitude, although such events may be rare.  Besides, axinovae is also an interesting phenomenon~\cite{Fox:2023aat}.

\section{A 5D model of axion and moduli}
 \label{sec:model}

Let us now discuss a concrete setup of moduli/axion which allows us to calculate the behavior of the axion 
during inflation.
We consider a 5-dimensional (5D) scenario where the moduli is the radius of the extra-dimension. 
\paragraph{Background metric}
To explain the setup, we first consider that the metric is a (non-dynamical) background, $i.e.$ the moduli is infinitely heavy.
The fifth direction is compactified on an orbifold $S^1/Z_2$, and the background metric is given by
\begin{align}
    ds^2=\eta_{\mu\nu}dx^\mu dx^\nu+dy^2\ ,
\end{align}
where $y$ is the extra space coordinate in the interval of $[0,L]$.
We assume that QCD is propagating in the bulk, so that the gauge coupling constant depends on $L$.
The gluon kinetic term in the bulk is given by 
\beq
{\cal L}_{5,kine} = -\frac{1}{4g_5^2} (G^a_{MN})^2
\eeq
with $g_5^2$ being a dimension $-1$ coupling, and $G_{MN}$ is the field strength
of the gluon field and we use $M,N, R, \cdots$ to denote the 5D space-time index.
By decomposing the 5D gluons as $(A^a_\mu,A^a_5)$, we take their
boundary conditions as $(N,N)$ for $A^a_\mu$ and $(D,D)$ for $A^a_5$,
where $N$ ($D$) denotes the Neumann (Dirichlet) boundary condition,
and the first and the second label correspond to the boundary
conditions for the two orbifold fixed points.
At low energy, the zero mode only appears for the 4D part, $A_\mu^a$, while the pseudo-scalar particle $A^a_5$ obtains a mass due to the boundary condition.
We now have the Standard Model as the low energy effective
theory with the QCD coupling as a function
of the radius.

The axion can be introduced in various ways.
One is to consider SUSY realization as in the next section
(\Sec{UVMODEL}), so that the axion is identified as the superpartner
of the moduli whose value determines $L$.
In this case we need to assume that there is no other
dynamics than QCD to give a mass to the axion.
We can also consider a scenario where the axion
arises from the gauge field in the extra-dimension 
via compactification.
By introducing a 5D $U(1)$ gauge field $(A_\mu,A_5)$, and taking
$(D,D)$ for $A_\mu$ and $(N,N)$ for $A_5$, we obtain a zero mode of
$A_5$, which is identified as the axion field. The axion-gluon
coupling is obtained by the 5D Chern-Simons term,
\begin{align}
\int d^4 x\int_0^{L}
dy \frac{b_{CS}}{64\pi^2} \epsilon^{MNRST} A_{M} G_{NR}^aG_{ST}^a\ ,      
\end{align}
where $b_{CS}$ denotes an integer.  Note that the CS term is invariant under the gauge
transformation of $e.g.$ $A_5\to A_5+\partial_5 \alpha(x,y)$ for $
\alpha(x,0)=\alpha(x,L)=0$.%
\footnote{These conditions of $\alpha$ are required to erase the boundary term.}

After the integration of the fifth direction, we obtain the effective
Lagrangian consistent with the one in Sec.\,\ref{sec:revisit},
 \begin{align}
\label{axion_effective}
S=\int d^4 x\left(
    -\frac{1}{4g_s^2[L]} (G^a_{\mu\nu})^2-\frac{1}{32\pi^2}\frac{a}{F_a[L]} G^a\tilde G^a-\frac{1}{2}(\partial_\mu a)^2
    \right)
\end{align}
where 
$g_s[L]$
is the QCD gauge coupling constant given as 
$1/g_s^2=L/g_5^2$
from the 5D gauge coupling $g_5$.
Importantly both the $g_s$ and $F_a$ are functions of the size of the
extra dimension, $L$, and thus change when the moduli has a different
value during inflation.
For instance, in the case where axion is originated from a 5D gauge
field, 
$F_a[L]$
is obtained 
as $F_a=\frac{1}{g_5 L^{1/2} b_{CS}}=\frac{1}{b_{CS}g_s L}$.

\paragraph{Dynamical metric and moduli}
Let us discuss the dynamics of the moduli, which appears as a degree of freedom from 5D gravity.\footnote{We call the radion the moduli in this paper as mentioned before.} 
The moduli field $\phi$ is introduced as a component 
in the 5D metric,
\begin{align}
    ds^2=\phi^{-1/3}g_{\mu\nu}dx^\mu dx^\nu+\phi^{2/3}dy^2\ .
\end{align}
Here, the vector fluctuation is omitted because it does not have zero
mode due to the orbifold projection. The physical size of the extra
dimension is given as $L\phi^{1/3}$. 
The five dimensional Einstein action reduces to the four dimensional
one such as
\begin{align}
M_5^3 \int d^5x \sqrt{G}R(G)
=M^2_P \int d^4 x \sqrt{g}
\left[
R(g)+\frac{1}{6}\frac{\partial_\mu \phi\partial_\nu\phi g^{\mu\nu}}{\phi^2}
\right] + \cdots\ ,
\end{align}
where $L M_5^3=M^2_P$.
There is a freedom to choose $L$ to an arbitrary value by rescaling
$\phi$. For usefulness, we took $L$ to obtain $\langle \phi\rangle=1$
at the present vacuum. Then, $L$ denotes the physical size of the
extra dimension in the current Universe.
During inflation, it may be
useful to go to the basis where the kinetic terms are canonically
normalized, 
\begin{align}
\phi=\exp\left(\sqrt{3}\frac{\hat \phi}{M_P}\right)\ .
\end{align}
In the basis with $\hat\phi$ ($\phi$), the present vacuum is given by
$\langle\hat\phi\rangle=0$ ($\langle\phi\rangle=1$).

 $g_s$ with arbitrary $\phi$ can be obtained as 
\begin{align}
\label{eq:gs_present}
\frac{1}{g_s^2}&=\frac{L\phi^{1/3}}{g_5^2}\ .
\end {align}
This can be different at different epoch of the cosmological history due to the moduli dynamics. 
The decay constant also depends on the moduli. 
In the case of the axion as 5D gauge boson, we have 
$
    F_a=\frac{1}{g_5 (L\phi^{1/3})^{1/2} b_{CS} \phi^{1/6}}=\frac{1}{g_s b_{CS} (L\phi^{1/3})\phi^{1/6}}\ ,
$
where $F_a$ scales by $\phi^{-1/6}$ for a fixed physical extra dimension length $L\langle\phi^{1/3}\rangle$.
As we will discuss below, we consider
$0<\langle\phi\rangle< 1$ ($\langle\hat\phi\rangle<0$ ) so that the
QCD gauge coupling constant during inflation becomes much larger than
the present value, and the QCD-induced potential plays important
role in the moduli stabilization.

\subsection{Moduli stabilization in the current Universe}
Now we are ready to discuss the moduli stabilization.
The moduli stabilization can be performed in more or less generic ways. 
This is because by integrating out various heavy fields, the moduli 1PI effective potential can be generically expanded as
\begin{align}
\label{eq:rad_vacuum}
    V = \phi^{-2/3} \left( \alpha L \phi^{1/3}+\beta +\frac{\eta}{L \phi^{1/3}} + \cdots
    \right) \ ,
\end{align}
where $\alpha,~\beta,~\eta$ are constants with mass dimension five, four, three respectively.
The overall factor of $\phi^{-2/3}$ is originated from
the transformation from the Jordan to the Einstein frame, and we have written the leading three terms
in the $1/(L \phi^{1/3})$ expansion. The expansion 
is justified as long as we have four dimensional effective theory in the weakly coupled regime.
The $\alpha$ term represents the 5D cosmological constant, and the $\beta$ term scales as the brane tension which has no volume dependence.
Terms including higher powers of $1/(L\phi^{1/3})$ are obtained in general, and we only include up to the $\eta$ term in the following discussion.
Indeed, the same form of the potential has been discussed in the LARGE volume scenario~\cite{Balasubramanian:2005zx} in which the third term in~\eqref{eq:rad_inflation} appears from the $\alpha'$ corrections to the K{\"a}hler potential.\footnote{We follow the notation of \cite{Ponton:2001hq} but we introduced a more dominant interaction and thus used $\eta$ rather than $\gamma$ to emphasize this difference. Although  the $\gamma \phi^{-2}/L^4$ term may be more dominant in certain UV realizations, we expect that during inflation $\phi<1$ and our $\eta$ term is more easily dominated. Introducing the $\gamma$ term will introduce more freedom for our scenario to work as we will see in \Sec{UVMODEL}.}
To find out the vacuum in the current Universe, the following two conditions are required,
\begin{align}
    &V=0\ ,~dV/d\phi=0\ .
    \label{eq:conditions}
\end{align}
These two conditions give us two relations among coefficients, 
\begin{align}
\beta&=-2\alpha L\phi^{1/3}\ ,~\eta=\alpha L^2\phi^{2/3}\ .
\end{align}
One can see that we need $|\alpha| \ll |\beta| \ll |\eta|$ in the
Planck unit to have $L \gg 1/M_P$.
The moduli mass around the present vacuum is obtained to be
\begin{align}
m^2_{\rm moduli}= \frac{2 \eta}{3 L\langle\phi\rangle M_P^2}\ ,
\end{align}
which means we need $\alpha >0$, $\beta < 0$ and $\eta >0$. Since
$\eta$ is at most $O(M_P^3)$, the moduli is generically light compared to
the Planck scale for $L \gg 1/M_P$. In SUSY models, $\eta$ is further
suppressed by the SUSY breaking scale.
For example, if we want to have a solution with 
\begin{align}
    L\sim 10^2/M_P\ ,~\alpha\sim 10^{-11}M_P^5\ ,
\end{align}
$\beta$ and $\gamma$ needs to be 
\begin{align}
 &\beta\sim -2\times 10^{-9}M_P^4\ ,~\eta\sim 10^{-7}M_P^3\ .
\end{align}
The moduli mass squared is calculated to be around $5\times 10^{-9} M_P$.
Note that we have taken $\langle\phi\rangle=1$ at the present vacuum so that $L$ represents the physical volume today.

It is important to note that we need delicate balance among terms in
the potential in order to obtain a vacuum of a weakly coupled theory
with vanishing cosmological constant. We need at least three terms to
satisfy the conditions in Eq.~\eqref{eq:conditions} at a finite value
of $L \phi^{1/3} \gg 1/M_P$. The leading two terms should be
suppressed somehow to be balanced with the third term. The moduli is
generically much lighter than $M_P$, and has a fragile potential under
the modification of the $V=0$ condition as well as the different
background field values.

\subsection{Moduli during inflation}

Given that the moduli potential in the current Universe
is under the delicate balance, it is quite natural
to assume that the moduli potential has a different
shape during inflation.
We parametrize the potential during inflation as
\begin{align}
\label{eq:rad_inflation}
    V=\alpha_{\rm inf} L \phi^{-1/3}+\beta_{\rm inf} \phi^{-2/3}+\frac{\eta_{\rm inf}}{L}\phi^{-1}\ ,
\end{align}
where the subscript $_{\rm inf}$, again, denotes the coefficients
during inflation, and the coefficients depend on time, $i.e.$ we
assume $\alpha_{\rm inf},\beta_{\rm inf},\eta_{\rm inf}$,
respectively, settle into $\alpha,\beta,\eta$ after the end of
inflation. 
In addition to the potential in \eqref{eq:rad_inflation}, the moduli potential receives
a contribution from non-perturbative QCD effects. We estimate the QCD confinement scale $\tilde\Lambda$, which can be understood as $\L_{\rm QCD}^{\rm inf}$ in Sec.\,\ref{sec:revisit}, by
\begin{align}
\tilde\Lambda=\phi^{-1/6}M_*\exp\left(-\frac{8\pi^2}{7 g_s^2(M_*)}\right) ,
\end{align}
where $M_*$
denotes the UV cutoff scale
where it satisfies $g_s^2(M_*) = g_5^2/L$ at the present vacuum ($\langle \phi \rangle = 1$)\footnote{The physical mass is given by $\phi^{-1/6}M_*$, $i.e.$ $\phi^{-1/6}$ factor appears when one considers the basis that the kinetic term of the 4D field is canonically normalized.}.
$\tilde\Lambda$ may be rewritten as
\begin{align}
\tilde\Lambda=\phi^{-1/6}\Lambda_{QCD}\exp\left(-\frac{8\pi^2}{7(g_5^2/L)}(e^{\frac{\hat\phi}{\sqrt{3}M_P}}-1)\right)
=\phi^{-1/6}\Lambda\ .
\end{align}
We include non-perturbative QCD effects by
\begin{align}
\label{eq:radion_potential_qcd}
    V=\alpha_{\rm inf} L \phi^{-1/3}+(\beta_{\rm inf}+\Lambda^4) \phi^{-2/3}+\frac{\eta_{\rm inf}}{L}\phi^{-1}\ ,
\end{align}
The moduli find its minimum during inflation with the
conditions of
\begin{align}
    &V=V_{\rm inf}=3M_P^2 H_{\rm inf}^2\ ,\\
    &dV_{\rm inf}/d\hat\phi=0\ ,
\end{align}
where $V_{\rm inf}$ denotes the total potential during inflation. 
These two conditions give us the relations
\begin{align}
\label{eq:betaI}
\beta_{\rm inf}&=-2\alpha_{\rm inf} L\phi_{\rm inf}^{1/3}+\frac{1}{7}\Lambda^4 (-7+\frac{32\phi_{\rm inf}^{1/3}\pi^2}{g_s^2})
+3\phi_{\rm inf}^{2/3}V_{\rm inf}\ ,\\
\label{eq:etaI}
\eta_{\rm inf}&=\alpha_{\rm inf} L^2\phi_{\rm inf}^{2/3}-\frac{32}{7g_s^2}L\phi_{\rm inf}^{2/3} \Lambda^4\pi^2-2V_{\rm inf}\phi_{\rm inf} L\ ,
\end{align}
where $\phi_{\rm inf}$ denotes the value of $\phi$ at the local potential minimum during inflation.%
\footnote{$\Lambda$ also depends on $\phi$.}

\begin{figure}
\centering
\hspace{1cm}
\includegraphics[width=0.6\linewidth]{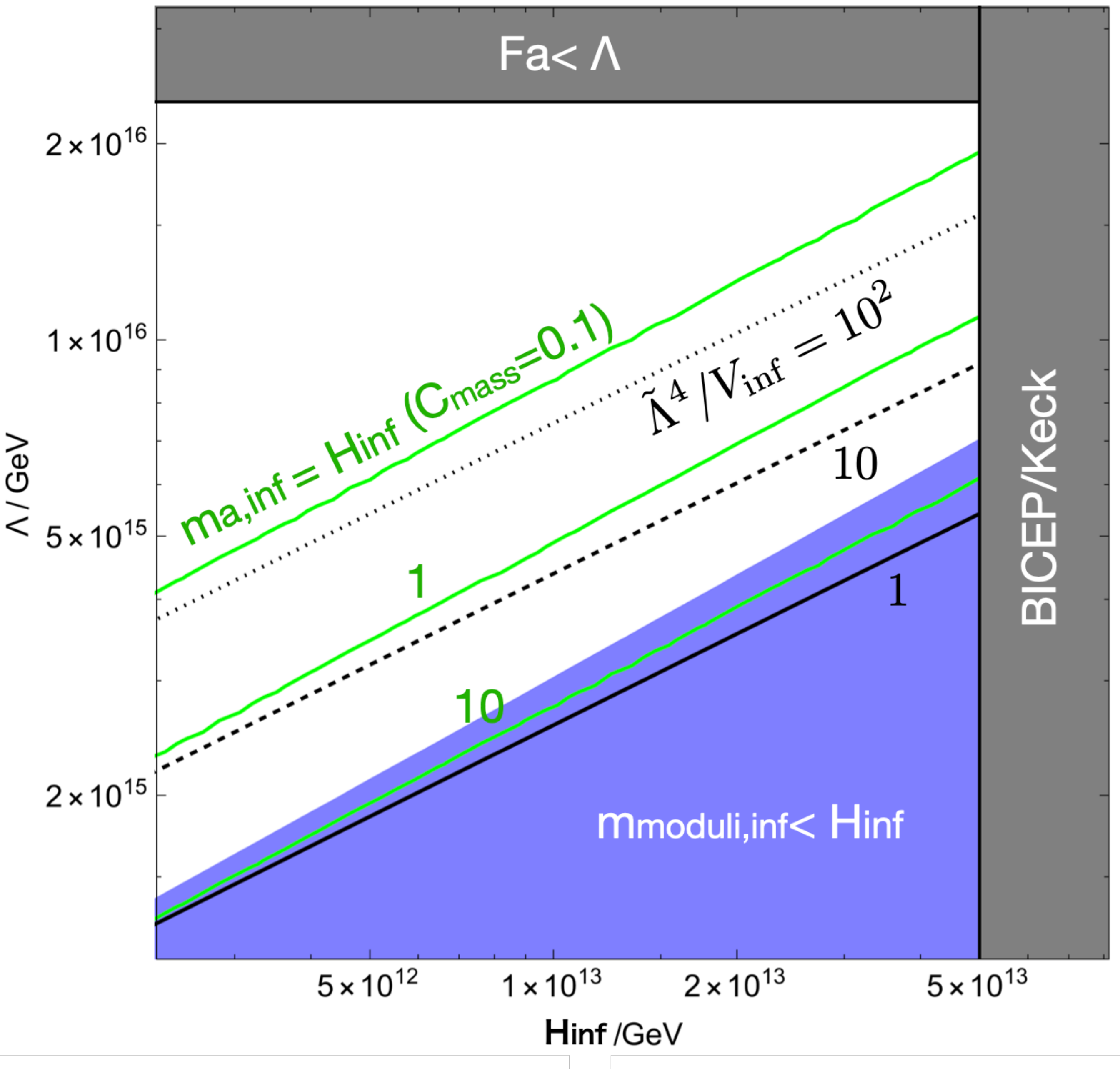}
   \vspace{0.2cm}
   \caption{
   The green solid lines denote $m_{a,\rm inf}=H_{\rm inf}$ for $C_{\rm mass}=0.1,~1,~10$, and we have $m_{a,\rm inf}<H_{\rm inf}$ in the lower region for each line.
   The blue shaded region denotes $m_{\rm moduli,inf}<H_{\rm inf}$. 
   The gray shaded region with horizontal boundary is given by 
$1/\tilde\Lambda<L\phi^{1/3}$.
   The black solid, dashed, dotted lines denote $\tilde \Lambda^4/V_{\rm inf}=1,~10,~10^2$, respectively.
   See the main text for details.}
\label{fig:H_L}
\end{figure}

%
In Fig.~\ref{fig:H_L}, 
we show the parameter region where the moduli and the axion can 
be heavy enough to be stabilized (non-colored region)
on the $H_{\rm inf} - \Lambda$ plane.
For small $\Lambda$, the axion and/or the moduli 
is lighter than $H_{\rm inf}$ so that the they cannot find their 
minimum during inflation (blue and thicker blue).
We introduce a parameter $C_{\rm mass}$ which represents the uncertainty associated 
with the axion mass during inflation,
for example, from the uncertainty of
the expectation value of the Higgs field during inflation.
We parameterize $m_a$ during inflation (cf.~Eq.~\eqref{eq:mainf}) as
\beq 
m_{a,\rm inf} = C_{\rm mass} \sqrt{y_u}\frac{\tilde\Lambda^2}{F_a} 
\eeq
with $C_{\rm mass}=[0.1-10]$.
In Fig.~\ref{fig:H_L}, the solid green lines denote $m_{a,\rm inf}=H_{\rm inf}$ for $C_{\rm mass}=0.1,~1,~10$ from upper left to lower right, respectively. In the blue shaded region, the condition $m_{\rm moduli,inf}>H_{\rm inf}$ is not satisfied.
Here, we took $M_*=M_P$, $L=10^2/M_P$, 
{$F_a=\frac{1}{(10^2/M_P)^{1/2} L^{1/2} \phi^{1/3}}$}
, and we also took $\alpha_{\rm inf}=-H_{\rm inf}^2$ for demonstration.
$\beta_{\rm inf}$ and $\eta_{\rm inf}$ are determined by the conditions in \eqref{eq:betaI} and \eqref{eq:etaI} for given $H_{\rm inf}$.
In the lower region for each green line, the axion mass is less than the Hubble parameter, $m_{a,\rm inf}<H_{\rm inf}$.
In the gray shaded region with horizontal boundary, we have 
{
$1/\tilde\Lambda<
L\phi^{1/3}$.
}
The black solid, dashed, dotted lines, respectively, denote $\tilde \Lambda^4/V_{\rm inf}=1,~10,~10^2$, which implies a cancellation between the potential terms of $1,~10\%,~1\%$, respectively. Above the lines we need more fine-tuning. 
We can see that the $m_{a,\rm inf} \sim H_{\rm inf}$ region indeed coincides with the less fine-tuned region.
Here $F_a \sim 5\times 10^{16}\,{\rm GeV}$ in the whole range of the figure.

\subsection{Scenario for axion dark matter}
We discuss the evolution of the moduli field value during inflation
and scenarios that the axion dark matter becomes the dominant one at
present. We assume that the local minimum of the moduli potential
continuously changes during inflation. In particular, we consider that
the moduli field value at the minimum gradually becomes large and
approaches the value at present ($\langle\phi\rangle=1$). To make the
moduli follow the local minimum, we require $m_{\rm moduli,inf}>H_{\rm inf}$ and 
\begin{align}
\frac{1}{m_{\rm moduli,inf}}\ll \left|\frac{\phi_{min}}{d\phi_{min}/dt}\right|\ ,
\end{align}
$i.e.$
\begin{align}
m_{\rm moduli,inf}\gg \left|\frac{d\phi_{min}/dt}{\phi_{min}}\right|\ ,
\end{align}
where $\phi_{min}$ denotes the moduli field value at the potential
local minimum~\cite{Linde:1996cx,Nakayama:2011wqa}. 
The moduli mass is
much larger than the inverse of the typical time scale of the change
of the potential minimum, and thus the moduli can quickly settle down
to the potential minimum. This can prevent the moduli from overshooting into the runaway regime.

As the moduli field value become larger, the gauge couplings becomes
smaller, and then the QCD confinement scale or the axion mass becomes
smaller. The axion mass has an exponential sensitivity to the change
of the moduli field value, and the axion mass can drastically change
compared to the moduli mass. If the axion mass becomes much smaller
than the Hubble scale during inflation, the axion starts to fluctuate,
and its fluctuation contributes to the current axion dark matter
abundance. 
This is the scenario discussed in Sec.\,\ref{sec:abundance}. We expect
the scenario to be probed in the future CMB experiments and axion DM
search, which may have a significant enhanced signal event rates with
a time dependence due to the minicluster/axionstar formation.

 We also note another possibility so that $m_{\rm a,inf} >H_{\rm inf}$ is
 satisfied until around the end of inflation. This is realized in a
 large region in the Fig.~\ref{fig:H_L} but it is with fine-tuning
 among the potential terms for the moduli stabilization. In this case,
 the current axion abundance is too small to explain the dominant DM.
 Even in this case, once we have a new source of CP violation in the
 axion potential during inflation. For instance, one can explain the
 large enough abundance by enhancing the small instanton
 effect~\cite{Kitano:2021fdl}.

\section{A SUSY model and suppressed CP violation}
\lac{UVMODEL}
The scenario in the previous sections assumes that
there is no CP violation that becomes important 
in the moduli or axion potentials during inflation.
The extra-dimension model is in this sense 
suitable as the space-time symmetry restricts
the CP violation to appear.  
Here we provide a possible UV model with many additional particles in the context of supergravity and show that the behaviors discussed so far as well as the CP safe nature hold. To this end, we consider 
$T$ is a moduli multiplet,
in which the previously discussed moduli field $\Re T$ is embedded. As a minimal possibility, we consider
the axion is $\Im T.$ 
We will consider that $\Re T$ is stabilized purely by K\"ahler potential contribution with SUSY breaking for satisfying the quality of the ``Peccei-Quinn symmetry", while 
during inflation, the QCD contribution becomes stronger, providing an alternative source of the stabilization of $\Re T.$

Let us consider the low energy SUSY theory in four dimension. The K\"{a}hler potential is given as
\begin{align}
\laq{seq2}
{\cal K}_4=-3M_P^2\log \frac{\(3 M_5^3 (T+T^\*)-F({Z_{\rm SB}},{Z_{\rm SB}}^\*)-\D K(T,T^\*)\)}{3 M_P^2}, \end{align}
The first term represents the Planck mass in terms of the extradimensional radius.  In addition, we introduce a SUSY breaking field $Z_{\rm SB}$ on a brane, and  $F$, the K\"ahlar potential, appears in the logarithm i.e. the sequestering formalism. $\D K$ is the correction of the K\"{a}hlar potential, and it may be from Casimir energy and/or leading $\alpha'$ correction motivated from the string theory. 
For later convenience, we consider the following super-potential, 
\beq
W= c+\D W(T)+\D Y({Z_{\rm SB}}).
\eeq
Here $c$ is the constant term that may arise from gaugino condensation on a brane. $\D W$ can appear from bulk QCD condensation, which is negligible in the present vacuum because it is from the ordinary QCD. 
However, it is non-negligible during inflation and plays an important role in stabilizing the moduli and axion at the era. This is the feature of our scenario and the new point of this model.
We also put a generic superpotential for ${Z_{\rm SB}}$, otherwise, the F-term would be vanishing. 
At the leading order of $\D K, F, \D W[T]$
 we get the supergravity potential
\begin{align}
V&\simeq -\frac{M_P^4 \left( M_5^{3} (c \Delta W^{\*}_{T^\*}+h.c.) +|c|^2 \Delta K_{T,T^\*}\)}{4M_5^{12} \Re(T)^2}+\frac{M_P^4|\D Y_{Z_{\rm SB}}|^2}{4M_5^{6} \Re(T)^2}+V_D
\end{align}
Here we neglect the higher order terms, e.g., $\frac{M_P^4   \Delta K |\D Y_{Z_{\rm SB}}|^2}{12M_5^{9} \Re{(T)}^3}$ in the present vacuum (see the following). 
We further imposed $F({Z_{\rm SB}},{Z_{\rm SB}}^\*)=\tilde{F}[{Z_{\rm SB}} {Z_{\rm SB}}^\*], \vev{{Z_{\rm SB}}}=\vev{{Z_{\rm SB}}^\*}=0, \Delta Y[0]=0, \partial_x\tilde{F}[x]|_{x=0}=1$ for simplicity of discussion. 
These conditions can be easily realized when the SUSY breaking field carries a charge under some symmetry. The last condition is the kinetic normalization for ${Z_{\rm SB}}$. 

As we have mentioned, \beq \D K\sim \frac{f(T+T^\*)}{16\pi^2 ( T+ T^\*)^2}
\eeq 
which can arise from Casimir energy e.g.~\cite{Arkani-Hamed:2004ymt}. Here $f(T+T^\*)$ represents a finite mass effect from the bulk matter. 

In addition, we may consider the D-term contribution by assuming the presence of a gauge field in the bulk. 
We get 
\beq
V_D=\sum_a \frac{g_4^2}{2} (\sum_i K_i T_a \phi_i)^2=\sum_a \frac{g_5^2  }{2 \Re T} (\sum_i K_i T_a \phi_i)^2 \equiv \frac{V_{ D,\rm coeff}}{\Re T}
\eeq
In particular, if this D-term is from an abelian gauge group (or asymptotically non-free gauge coupling) we do not need to care about the non-perturbative effect to the moduli potential.

The kinetic term of the four dimensional gravity in this formalism is already normalized,
but the kinetic term of $T$ in this formalism is 
\beq
{\cal L}_{kin}=3 M_P^2 \frac{|\partial T|^2}{\(T+T^*\)^2}\supset 3 M_P^2 \frac{(\partial R)^2}{4 R^2}
\eeq
By noting the kinetic normalization condition, $\sqrt{\frac{3}{4}}\log\Re (T)=\sqrt{\frac{1}{12}}\log \f\to \Re T \propto \f^{1/3} $, we get the potential in terms of $\f$ as
\beq
V(\phi \sim 1)\simeq \frac{ \alpha }{\f^{1/3}}+\frac{\b}{\f^{2/3}}+ \frac{ \g(\phi)}{\f^{2}}.
\eeq
$\a$ originating from $V_{D,\rm coeff}$ is vanishing if the VEV of $\phi_i$ is absent. Here, we take the redundant parameter $L$=1 for simplicity of discussion.
At the loop level, the bulk gauge coupling $g_5^2/\Re[T]$ enters into the running of the K\"{a}hler potential. Thus in general we have contribution scaling with $1/(\Re T)^3$ from loop corrections from either $\a$ and $\beta$ terms,
\beq
\d V = \eta \phi^{-1}
\eeq
Since it is a loop suppressed effect we do not consider it in the present Universe  but it will be more important than the $\gamma$ term when $\Re T$ is small during inflation.
It is known that in the absence of $\L$, $\eta$, and $V_{D,\rm coeff}$, which is assumed in the present Universe, that the moduli can be stabilized~\cite{Luty:2002hj}. Namely, the stabilization can be successful only with the presence of $\beta$ and $\gamma(\phi)$ term with certain bulk fields, thanks to the non-trivial $\phi$ dependence in the Casimir energy.

In this case, for the vanishing cosmological constant, 
 $|\D Y_{Z_{\rm SB}}|^2$ is suppressed compared with $|c|^2$ since $M_5^{-2}\D K$ is both loop- and $M_5/M_P$ suppressed, 
 \beq\laq{stabilization}
\vev{| Y_{Z_{\rm SB}}|^2}\simeq \vev{\frac{2|c|^2\Re (T)\Delta K_{T,T^\*}}{M_5^3 M_{P}^2}}\sim \frac{\vev{|c|^2} M_5^6 }{16\pi^2 M_P^8}=\frac{1}{16\pi^2} \frac{m^2_{3/2} M_5^6}{M_P^4}
 \eeq
 We put $\vev{}$ for $F_{Z_{\rm SB}}$ to emphasize that it is the condition for cancelling the cosmological constant in the vacuum.
This is different from ordinary supergravity because the gravitino mass is heavier than the SUSY breaking, which is a well known feature of the (almost) no-scale supergravity~\cite{Cremmer:1983bf,Lahanas:1986uc,Luty:2002hj}. 

During inflation, on the other hand, many fields, including $T$, can be away from the potential minimum. Since we consider a small $\Re T$, then the loop suppressed $\O(1/\f)$, $\L$, and $\a$-term from a D-term contribution can be important depending on the inflation model building. Thus by using the small $\Re T$ expansion, we have the dominant terms
\beq
V(\f\ll 1)\sim \frac{\a}{\f^{1/3}}+ \frac{(\b+\L(\f))}{\f^{2/3}}+ \frac{\eta}{\f}.
\eeq
This is the generic form that we have used. As we have mentioned, we can also take account of the $\alpha'$ correction from string theory, e.g. Refs.\,\cite{Cicoli:2008va,Kobayashi:2018pbp}, to get the last term. One can replace $T+T^\*$ by $T+T^\*+ \x$ and expand the potential in terms of $\x$. 

Another interesting feature is the CP-safety in the colored sector. Indeed, it is also known that the soft mass spectrum can be CP-conserving if $T$ dominantly mediates the SUSY breaking (with R-symmetry)~\cite{Choi:1993yd, Choi:2006za,Iwamoto:2014ywa}. Thus a good quality of the CP symmetry is guaranteed in our scenario, if during and after inflation the moduli F-term has the dominant SUSY breaking contribution to the particles that are relevant to the QCD.\footnote{
We note that $6$-flavor SUSY QCD may be conformal in the low energy EFT, although the Higgs Yukawa coupling and gauge coupling may give relevant contributions. During inflation, the colored particles can decouple either  due to some large expectation values of  a combination of charged fields respecting SUSY or due to the decoupling of heavy SUSY partners. Thanks to the CP-safety, the additional CP phases do not shift the minimum of the strong CP phase in a CP violating way. In some CP-conserving case, the phase may be shifted by $\pi$ which is the exception of the argument~\cite{Daido:2017wwb,Co:2018mho, Takahashi:2019pqf}. We do not consider this possibility in this paper. 
} 
In this scenario, the shift of the axion minimum during and after inflation is naturally suppressed. Thus during inflation the axion is stabilized very close to the CP-conserving minimum.

We consider this parity safety to be more generic.
This is due to the fact that the parity in 4D effective theory is a part of the rotational invariance in the 5D theory and thus conserves.
At this stage, one can assign parities for moduli and axion as even and odd, respectively.
The orbifold projection would not disturb this assignment 
as both fields are not projected out.
The parity violating contribution to the moduli/axion potential can be in principle generated by the boundary effects, which, however, are suppressed by the 
volume of the extra dimension.
 Since the gluon, axion, and the moduli live in the bulk, the parity-violating couplings of moduli and gluon are naturally suppressed by the 5th dimensional volume and thus our setup can naturally preserve the feature needed for suppressing the axion misalignment angle during inflation.\footnote{A caveat may arise from the fact that the scale of $1/g_5^2$ is close to the extra-dimensional scale to enhance the inflationary QCD. We implicitly assumed that we do not have the CP violating higher dimensional terms with a scale around or smaller than this, i.e. the QCD has one of the strongest couplings. Since a less than $O(0.1)$ CP violating contribution to the misalignment angle during inflation is still consistent with our scenario, we expect this is not a very strong assumption.}

\section{Conclusions and discussion}
\label{sec:conclusion}

The axion abundance is quadratically sensitive to the initial
amplitude. In the case where the axion exists as an
effective degree of freedom before the inflation era, it is commonly
assumed that the initial value is set by a random choice, and also the
model is subject to the constraint from the isocurvature perturbation
due to the fluctuations of the axion field during inflation.

This discussion is justified under the assumption that the inflation
dynamics would not affect the axion potential.
We show, in this paper, the axion can actually be heavier than the Hubble
parameter during the inflation so that the axion field value and the
fluctuations are stabilized. For example, we discussed a string-theory
motivated scenario of our spacetime, where there is an extra dimension
as small as the size of the string/GUT scale.
As it is usually the case, the volume of the extra dimension is
delicately tuned in order to realize our almost flat spacetime by
balancing various potential terms such as the Casimir energy and the
scale/volume dependent coupling constants.
The inflation dynamics, i.e., the extra energy source to bring the
flat spacetime to de Sitter space, generically interferes with this
delicate tuning during inflation, and modifies the size of the extra
dimension.
It is possible to make the size smaller by a factor of $O(10)$, which
in turn modifies the coupling constants in the Standard Model. If the
gluon fields propagate into the extra dimension, the QCD coupling is
enhanced, and a very large axion mass can be realized.
Since the source of the axion potential is dominated by the QCD
dynamics, the minimum of the potential during the inflation is the
same as the current one as long as there is no new CP violating
contribution to the axion potential during inflation. We discussed the
possible sources of the new CP phases, and found that the scenario of
the volume modulus naturally avoids such phases.

This scenario gives a unified picture of the axion and the spacetime.
The size of the extra dimension and the axion decay constant can both
be of the order of the GUT/string scale. One can have a natural
picture of identifying the axion as a component of a gauge field which
have the Chern-Simons coupling to QCD in the bulk for the super partner of the moduli. It is interesting
that the abundance of dark matter may be telling us what happened at
the very beginning of our Universe as it is sensitive to the initial condition. 

So far, we have considered the moduli and inflaton are different degrees of freedom. A simple unified possibility is, however, that the moduli is inflaton. In more detail, we have assumed a particular motion of the moduli, $i.e.$ the field value of the moduli is continuously changing during inflation and it is stabilized soon after the end of inflation. Such a motion indeed evokes inflaton.

\section*{Acknowledgements}
M.S. and R.K. would like to thank Fermilab theory division for their hospitality, and especially would like to thank Paddy Fox and Bogdan Dobrescu for useful discussions.
This work was supported by JSPS KAKENHI Grant Numbers JP22J00537 (M.S.),
JP22K21350 (R.K.),
JP21H01086 (R.K.), 
JP19H00689 (R.K.), 
20H05851 (W.Y.), 21K20364 (W.Y.), 22K14029 (W.Y.), and 22H01215 (W.Y.).

\bibliographystyle{utphys}
\bibliography{bib}

\end{document}